\documentclass{article}
\usepackage{LaThuileFPSpro}
\usepackage{epsfig}

\newcommand{\ttbar}{ {\rm t} \bar{{\rm t}} }
\newcommand{\ipb}{ {\rm pb}^{-1} }
\newcommand{\ifb}{ {\rm fb}^{-1} }
\newcommand{\ppbar}{ {\rm p} \bar{{\rm p}} }
\newcommand{\qqbar}{ {\rm q} \bar{{\rm q}} }
\newcommand{\et}{{\rm E}_{\scriptscriptstyle\rm T}}
\newcommand{\pt}{{\rm p}_{\scriptscriptstyle\rm T}}
\newcommand{\met}{\mbox{$\protect \raisebox{.3ex}{$\not$}\et$}}
\newcommand{\bbbar}{ {\rm b} \bar{{\rm b}} }
\newcommand{\ccbar}{ {\rm c} \bar{{\rm c}} }

\begin{document}

\title{ TOP MASS MEASUREMENTS AT THE TEVATRON RUN~II 
      }

\author{
   Gueorgui V.~Velev  \\
   e-mail:~velev@fnal.gov \\
   {\em  Fermi National Accelerator Laboratory, Batavia, IL 60510} \\
   {\em for}\\
   {\em the CDF and D\O~ collaborations}
}
\maketitle

\baselineskip=11.6pt

\begin{abstract}
The latest top quark mass measurements by  the CDF 
and  D\O~ experiments are presented here. The mass has been determined 
in the dilepton ($\ttbar \rightarrow  e\mu,ee,\mu\mu$ + jets +$\met$) 
and  lepton plus jets ($\ttbar \rightarrow e$ or $\mu$ + jets +$\met$)
final states. The most accurate  single result from lepton plus jets channel 
is  173.5$^{+3.7}_{-3.6}$(stat. + Jet Energy Scale Systematic)$\pm1.3$(syst.)
~GeV/c$^{2}$, 
which is better than the combined CDF and D\O~ Run~I  average.  
A preliminary and unofficial average of the best experimental Run~II results gives 
$M_{top}$ = $172.7\pm3.5$~GeV/c${^2}$.
\end{abstract}



%
\newpage
\section{Introduction}

\hspace*{0.8cm} 
Since the first evidence in 1994\cite{evidence}  and the discovery 
of the top quark in 1995\cite{cdf-discovery}\cite{d0-discovery}, the CDF and  D\O~
Collaborations   invested a lot of work to determine   its  properties, specially
the value of its mass, which is a fundamental parameter of the Standard Model (SM). The   
ongoing Run~II, with the upgraded Fermilab Tevatron collider and CDF  and D\O~ detectors, 
gives  new possibilities  for a precise measurement of the top mass.
Due to its large mass, corresponding to a Yukawa coupling of order unity,
one may suspect that 
the top quark may have a special role in the electroweak symmetry breaking. In addition,
due to its significant contribution to high order radiative  corrections
of a number of electroweak observables, a precise 
measurement
of the top quark mass
provides  a tighter constrain on the Higgs mass\cite{talk-in-this-proceeding}.       
  
This paper reports on the latest CDF and  D\O~
top quark mass results which are based on  about 
318~$\ipb$ (CDF) and 219~$\ipb$ (D\O~) of data from
the first two years of the Tevatron  Run~II (2002 to 2004). 
Another paper, presented on 
this conference\cite{kinematics}, summarized the top quark  
kinematics properties including its 
recently measured  production cross section at  the center of mass energy 
of $\sqrt{s}$=1.96 TeV, 

At this Tevatron energy,
top quarks are produced generally in  pairs from the processes
$\qqbar \rightarrow \ttbar$ (in $\sim$85\% 
of the cases) and  $gg \rightarrow \ttbar$ (in $\sim$15\%
of the cases).
Top can be produced as a single quark  by
electroweak interactions, by W-gluon fusion or  virtual
W$^{*}$ production in the s-channel\cite{single-top}, but with a smaller cross section. 
At this time, no signal has been observed from single top
processes and they are not expected to be utilized for a precise mass 
measurement.

In the Standard Model the  branching ratio of the decay $t
\rightarrow bW$ is nearly 100\%. 
When a $\ttbar$ pair is produced,  each of the W-bosons can  decay into either 
a charged 
lepton and a neutrino (branching ratio of 1/9 for each lepton family) or
into a $\qqbar^{'}$ quark pair  (branching ratio of 2/3). This allows  us 
to classify the final states as:
\begin{itemize}
\item Dilepton final state, when both W's from the $\ttbar$ pair 
decay leptonically. This state is characterized by  two  high 
P$_{T}$  charged leptons, 
two jets from $\bbbar$ quarks \footnote{Errors in jet reconstruction and gluon 
radiation in the event may make the observed number of jets smaller or larger. This 
statement is  valid for all final states.} 
and significant missing transverse energy ($\met$) from the neutrinos.    
\item Lepton plus jets final state, when one W boson decays leptonically
and the other one hadronically. This state contains one high P$_{T}$ 
charged lepton, 
four jets and significant $\met$. 
\item All hadronic final state, when  both W's decay hadronically.
This state 
is characterized by  8 jets, two of which are from $b$-quarks.
\end{itemize}

In Run~I,  CDF and D\O~ used all of these signatures for  the top  mass measurements.
At this time, the Run~II top mass analyses  from the all-hadronic  channel are still in 
progress and will not be reported in this paper.    

\begin{figure}[t]
  \vspace{9.0cm}
 
  \includegraphics{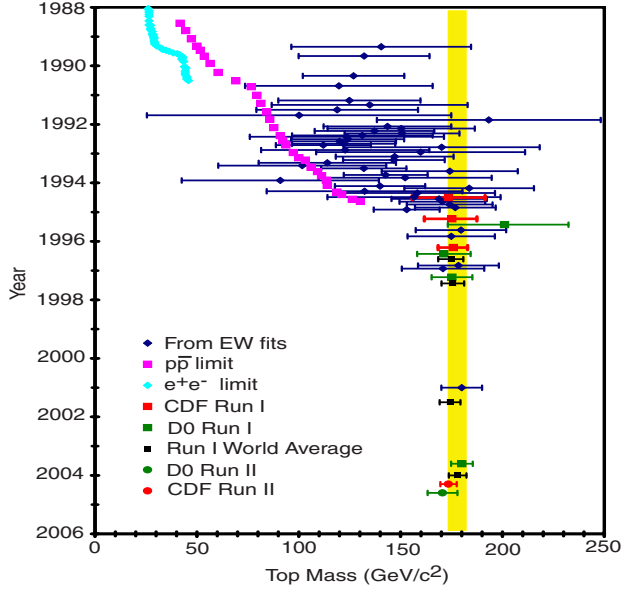}
  \caption{ \it
The summary of the top mass ``evaluations'' and direct measurements
 versus time. See the text for 
explanation of the points. 
   \label{fig:topmasshistory} }
\end{figure}

Figure~\ref{fig:topmasshistory} shows how  
our knowledge on the top quark mass improved with time. 
The diamonds represent the  indirect determinations 
from  fits to the  electroweak observables\cite{quig}. The curves  in the 
upper left corner of the figure represent the limits from direct searches in 
$e^{+}e^{-}$ and $\ppbar$  machines. The Run~I CDF and  D\O~ results  are  presented 
with 
squares. The filled  circles are the new Tevatron Run~II  results. The band 
represents 
the average Run~I top mass results. One observes that the 
recent 
Tevatron  results have equal or better accuracy than the Run~I world average.    
 
\section{Top kinematics and mass reconstruction methods}
\hspace*{0.8cm} The  kinematics of the events in the lepton plus jets final state 
is over-constrained. In this channel, the number of measured quantities and the 
number of applicable energy-conservation equations, from five production 
and decay vertices, are larger than the number of non-measured kinematical event
parameters.
This feature allows for a complete  reconstruction of the four-momenta of the  
final state particles in
the event, for example using  the two constrain kinematical fit 
(2CF), and for a reconstruction of the top mass event-by-event. 
In this type of analysis
there is  an ambiguity in how to assign the  four leading jets to the 
two b quarks  and two light quarks coming from the $\ttbar$ system. 
If none of these jets is tagged by b taggers 
\footnote{CDF and D\O~ use taggers based on either 
displaced vertices (Secondary Vertex Tagging, SVX, for example see\cite{acosta_hep}) or on low 
$P_{T}$  electrons or muons  from the $b$-quark semileptonic decays 
(Soft Lepton Tagging, SLT\cite{kestenbaum_PhDthesis})},  
there are 12 different ways of assigning jets to the 4 partons. 
Combining with the ambiguity from solving of a 
quadratic equation on $P_{z}^{\nu}$, there are  24 different 
values of $m_{top}$ returned by the fit.
The combinations are reduced to 12 or to 4 if one or two of the  jets are selected 
from b-taggers.      

The variety of the CDF and  D\O~ analyses in the lepton plus jet sample is 
relatively 
large. However, all  analyses  can be separated  in three major categories:
\begin{enumerate}
\item Template type analyses, where the reconstructed mass distribution from the 
data is  compared with expected distributions from Monte Carlo generated 
signal (mass-dependent) and background. In these types of analyses, all events are weighted 
equally. By doing so one neglects the additional  information coming 
from a different mass resolution in single events.
In addition no use is made of possible information 
from the 
dynamic of the  process which can be assumed to be known.  
Typical examples for this type of analysis are the 
CDF and D\O~ Run~I\cite{CDFrun1PRD}\cite{D01998top} and  the latest, most accurate CDF Run~II 
lepton plus jets analysis, described in Section~3.1.
  
\item Matrix Element type analyses, originally  proposed by Dalitz and Goldstein\cite{dalitzgoldst}  and independently by Kondo\cite{kondo}. 
These  methods calculate the posterior probability, 
given the known production cross section,
for every event with measured 
kinematic properties, to originate  from a $\ppbar \rightarrow \ttbar$ process.
A typical example is  the D\O~ Matrix Element analysis which is  the base of 
the  best Run~I top mass measurement\cite{nature}. In Run~II CDF uses a method, 
proposed by K.Kondo\cite{kondo},  called Dynamic 
Likelihood Analysis, which  differs in the way that
the  normalization of  the differential cross section is  performed.

\item A mixture between methods 1) and 2). For example  the $\ttbar$ event 
is reconstructed 
using the kinematic algorithms similar to the template analyses but an event-by-event 
probability  of each  kinematic reconstruction
is exploited as a weight (for example  $exp(-\frac{\chi2}{2})$). 
A typical example of this category is the  D\O~ 
Ideogram Analysis\cite{D0ideogram}.           
\end{enumerate}

\section{Lepton plus jet channel}
\subsection{CDF  result}
\hspace*{0.8cm} For the time being the most accurate top quark mass measurement 
comes from the lepton 
plus jets channel. This channel combines  the benefits of 
good signal to
background ratio, the possibility to reconstruct the top quark mass event 
by event with a relatively small combinatorial effect, and a large branching 
fraction.  
In brief, we discuss below  the main  selection criteria for this  channel.

Lepton plus jets events have the signature 
$ \ppbar \rightarrow \ttbar X \rightarrow \ell \nu b q \bar{q}^{'}
\bar{b} X$. The
characteristics of this final state begins with  the identification  of 
one isolated central high energy lepton ($e$ with $\et>$ 20 GeV or 
$\mu$ with $P_{T}> $20 GeV) and $|\eta| <$~1 \footnote{ 
A complete description of the lepton selection, including all cuts used, can be 
found 
elsewhere\cite{17cdftopprd}}. 

Assuming that  the  lepton is  coming from W boson decay, a companion neutrino 
should 
exist. This  would spoil the balance of the  observed momentum in the transverse 
plane.
The missing transverse energy ($\met$) is  constructed by adding the calorimeter 
energy vectors 
in the plane transverse to the beam.
The calorimeter clusters identified as jets are  
corrected for detector response and  for multiple $\ppbar$ interactions. 
In muon events the $\met$ is computed using the muon momentum measured 
by the track instead of the muon calorimeter signal.

In order to fully reconstruct the $\ttbar$ system, at least four central jets 
$|\eta| 
\leq $~2 are required in the system. The SVX tagging algorithm is run over 
the leading jets ($E_{T} > $15 GeV): some of them may be  
identified as b-jets. To obtain maximum 
statistical benefit from the  event sample it is helpful to decompose it 
into several classes of events which are expected to have different 
signal-to-background 
ratios and top mass resolutions. The Monte Carlo studies showed that an
optimal partitioning is obtained  splitting the sample  into four statistically independent
categories: events with double SVX tags (2SVX), events with single SVX tag 
and tight 
forth jet cut (four jets with $E_{T} > $15 GeV, 1SXVT), events with single SVX tag and 
loose forth 
jet cut (15  $>E_{T} > $8 GeV for the 4$^{th}$ jet, 1SXVL) and  finally events without
tags (0-tag).  Since the  last sample has a high  background 
contamination  compared to SVX tagged ones, 
an additional optimization of the jet $E_{T} $ cuts ( $E_{T}> $ 21~GeV) 
was performed. A total 
of 165 $\ttbar$ 
candidates were selected from 318~$\ipb$ of data.  

The dominant backgrounds  in  all   samples are
direct W plus multijet production, including heavy  flavour production,  
and QCD multijet events where one jet is misidentified as a
lepton. Additional small backgrounds are due to WW/WZ and single top production.   
The amount and composition of the background depends on the sample. 
In the case of the 2SVX sample,  
the $W\bbbar+W\ccbar+Wc$
background dominates ($\sim$60\%) while in the case of the 0-tag sample, the W plus 
light quark  
production is responsible for $\sim$75\% of   background.   
The information  for these four subsamples, including the dominant type  
and background fraction,
is  summarized in Table~\ref{t-sample}.

\begin{table}[t]
  \centering
\caption{\it In the first four columns from left to right: lepton plus jets subsamples used in the top quark mass
analysis, number of events in each sample,
S/B ratio, and a summary of the  jet energy cut selection are presented. 
A total of 165 
$\ttbar$ candidates 
were selected. The last column summarizes the background fractions in  
\% from  $W$ + light quark, 
$W\bbbar+W\ccbar+Wc$ and QCD multijet events (left to right).}
  \vskip 0.1 in
  \begin{tabular}{|c|c|c|c|c|} \hline

  Data                     & Number & S/B    &  Jet $E_{T} $ cuts (GeV)  & Bckg. type and  \\
  Subsample                & of events &        &   jets 1-3  (4$^th$ jet ) &  fraction in  \%  \\ 
     
    \hline
    \hline 
2SVX                              & 25       & 10.6/1 &    $E_{T} > $15 ( $E_{T} > $15) &  21/59/10\\ 
1SVXT                             & 63       & 3.7/1  &    $E_{T} > $15 ( $E_{T} > $15) &   17/38/22\\
1SVXL                             & 33       & 1.1/1  &    $E_{T} > $15 ( 15  $>E_{T} > $8) & 29/48/14\\ 
0-tag                             & 44       & 0.9/1   &    $E_{T} > $21 ( $E_{T} > $21) &  75/3/20 \\ 

    \hline
  \end{tabular}
  \label{t-sample}  
\end{table}


Each event, either  from data or MC samples,   
is fitted to  the hypothesis  $ \ttbar X \rightarrow \ell \nu b q \bar{q}^{'}
\bar{b} X$. We use four  kinematic constraints, as a consequence of the assumed lepton 
plus jets
event structure ($M_{\ell\nu}$ = $M_{jj}$ = $M_{W}$ and 
$M_{\ell\nu b}$ = $M_{jjb}$ = $M_{top}$). 
The fitting procedure runs over all 
possible 24 combinations of assigning the four leading jets to
the  $b$, $\bar{b}$ and  $W  \rightarrow q \bar{q}^{'}$ partons (the order of the pair 
assigned to the W is irrelevant). 
If one or two of the four leading jets are tagged as a $b$-jets, they are 
assigned to the $b$-partons and the number of explored
combinations is correspondingly smaller.   
All solutions  with  $\chi^{2} < $ 9  (cut optimized on the MC studies) are 
accepted. The top mass value corresponding to the 
combination with the minimum  $\chi^{2}$ is picked as the 
mass value indicated by the event.

The events from the MC samples are used to produce probability density  
distributions  or so-called templates. 
In case of the signal MC, these distributions are parameterized  as a function  
of reconstructed  
and input top masses. On the other hand, the background probability density 
distributions are parameterized 
only as a function of the reconstructed mass. The likelihood of each  
subsample  uses 
the parameterized signal and background probability density functions (p.d.f) to 
evaluate the dependence of 
the likelihood on the input top mass. 

To reduce the dominant systematic error coming from jet energy scale (JES)\cite{CDFrun1PRD}  the
latest  CDF template analysis exploits the fact  that  the global JES scale 
can be 
determined   from the decay $ W \rightarrow  q \bar{q}^{'}$. MC studies have 
shown that  this 
technique provides a 22\% reduction in the JES uncertainty. Similar template 
distributions as  for the kinematically 
reconstructed top mass are built for the dijet mass, with the exceptions of 
removing the 
$\chi^{2} < $ 9 requirement, exploiting all the possible jet-to-parton 
assignments in the event.
Examples of  top and dijet reconstructed masses p.d.fs are shown on   
Figure~2
left and right 
respectively.

\begin{figure}[t]
  \vspace{9.0cm}
  \includegraphics{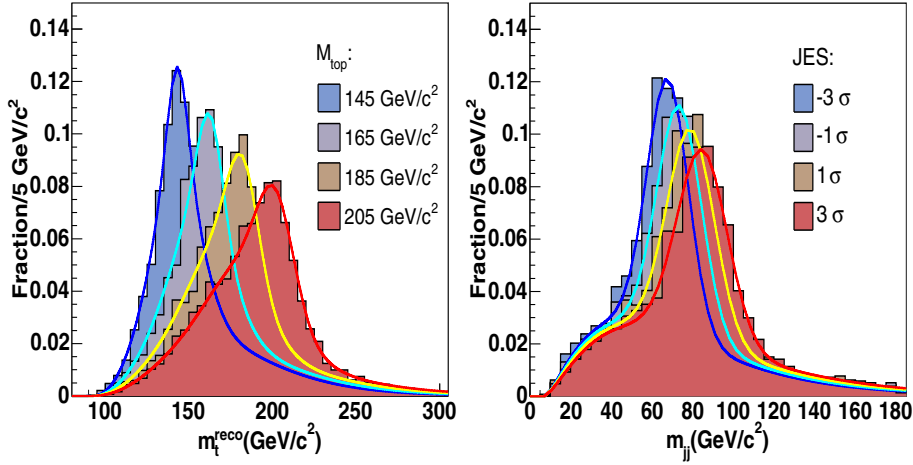}
  \caption{\it
 Left: reconstructed  top mass p.d.fs for input top  masses from 145 to 
205 GeV/$c^{2}$ and JES = 0. Right:      
reconstructed  mass p.d.fs of the dijet system attributed to the W
for different JES values in the range 
of $-3\sigma~-~+3\sigma$ and input top
mass of 180 GeV/$c^{2}$. 
    \label{fig:pdfs} }
\end{figure}


The reconstructed top and W dijet mass values for every data event are 
simultaneously compared to the p.d.fs from 
signal and background sources performing an unbinned likelihood fit. The 
fit finds a maximum likelihood value 
according to: the expected numbers of signal and 
background events in each subsample, 
the JES and the true top quark mass ($M_{top}$). Only  $M_{top}$ is a free 
parameter  in the  likelihood fit, the other 
are constrained within their uncertainties. For each subsample, 
the likelihood has the following form:

\begin{equation}
\mathcal{L}=\mathcal{L}_{shape}^{M_{top}} \cdot \mathcal{L}_{shape}^{m_{jj}} \cdot \mathcal{L}_{counting} \cdot \mathcal{L}_{bg}.
\label{lhood}
\end{equation}
    
In~(\ref{lhood}),  the main information on the top quark mass 
is hidden in the term $\mathcal{L}_{shape}^{M_{top}}$. It gives 
the probability for an event with reconstructed top mass 
${m_{top}^{rec}}$ to come from  true top mass $M_{top}$.
All other terms constrain the JES ($\mathcal{L}_{shape}^{m_{jj}}$), the number of 
observed events ($\mathcal{L}_{counting}$) and number
of expected backgrounds in the subsample  ($\mathcal{L}_{bg}$) and help  to 
reduce  
the statistical and systematical uncertainties returned by the fit.

The systematic uncertainties are summarized in Table~\ref{tab:syst}. For each 
systematic source, the relevant 
parameters are varied by $\pm1\sigma$ in the $\ttbar$ MC sample with $M_{top}$=178 
GeV/c$^{2}$ and 
sets of fake events are generated. These fake events are reconstructed in the 
same way as normal events. This 
procedure is  called  ``pseudoexperiments'' (PE). It propagates the $\pm1\sigma$ 
effects to a shift in the  
top mass 
relative to the result from the nominal sample.

The reconstructed top masses in the four subsamples with overlaid best fit 
for the signal and background 
MC expectation   
are shown in Figure~3.
 The  combined fit for  all 
lepton plus jets events 
returned $M_{top}$ = 
173.5$^{+3.7}_{-3.6}$(stat.+JES) $\pm1.3$(syst.)~GeV/c$^{2}$ and 
JES = -0.10$^{+0.89}_{-0.91}$(stat.+syst.). 
This is the most precise single  measurement available to date, 
better than the average Run~I  result.

\begin{table}[t]
  \centering
\caption{\it The systematic uncertainties  in the
CDF lepton plus jets top quark mass measurement.
    }
  \vskip 0.1 in
  \begin{tabular}{|l|c|} \hline
Source 	    		&     $\Delta M_{top}$        \\
                        & (GeV/c$^{2}$)           \\
    \hline
    \hline
  b-jets modeling	                       &    0.6  \\
  Method                                       &    0.5  \\
  Initial state radiation                      &    0.4  \\
  Final state radiation                        &    0.6  \\
  Shape of background spectrum 	        &    0.8  \\
  $b$-tag  bias		                &    0.1  \\
  Parton distribution functions         &    0.3  \\
  Monte Carlo generators                &    0.2  \\
  MC statistics 	                &    0.3  \\
\hline
Total			                &    1.5  \\
    \hline
  \end{tabular}
  \label{tab:syst}
\end{table}



\begin{figure}[t]
  \vspace{9.0cm}
  \includegraphics{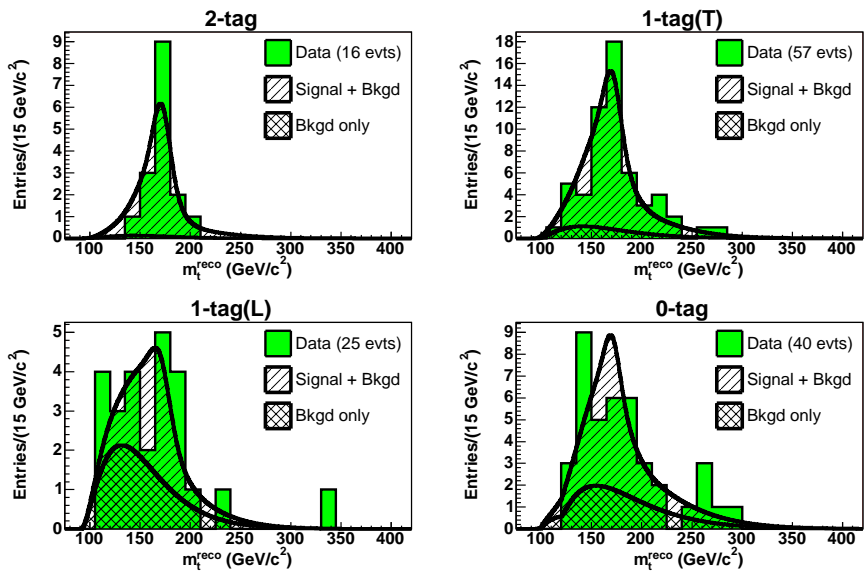}
  \caption{\it
The reconstructed mass distribution (histogram)  
for each  lepton plus jets CDF  subsamples is 
overlaid with the result of the likelihood fit
(signal+background, hatched area). The cross hatched area represents only
the background.
    \label{fig:topres} }
\end{figure}


\subsection{ D\O~  Result}
\hspace*{0.8cm} 
D\O~has  measured the top quark mass in the lepton plus jets channel as well. 
The utilized data set corresponds to an integrated luminosity of approximately 
229~$\ipb$
collected between April 2002 and March 2004. 

The event selection criteria are similar to those used in CDF. As a first 
selection step,
an identification of the a high $P_{T}$ isolated electron or muon accompanied by 
substantial large 
$\met >$ 20~GeV is required. The isolated electron (muon) candidate should have
  $P_{T}>$20~GeV,
satisfy a pseudo-rapidity cut of $|\eta|<$1.1 ($|\eta|<$2.0) and tight quality 
conditions. These initial selections
provide  the  data sample.

Two separated analyses, b-tagged and topological, are performed on this  sample. 
In the b-tagged analysis, to 
reconstruct the top mass the  events are additionally selected to have at least 
4 jets with  $P_{T}>$ 15~GeV and $|\eta|<$2.5.
A further requirement of identification of one or more jets as $b$-jets is 
made. A jet is b-tagged  
based on the reconstruction of the secondary verticies using 
the charged particle tracks associated with it.
49 (29) $e$+jets ($\mu$ + jets)   b-tagged events survive  all  cuts and  are 
kinematically fitted to the $\ttbar$ hypothesis. 
In 42 (27)
electron  (muon) events the kinematic fit converged in a configuration where 
the lowest $\chi^{2}$ solution is  consistent  with 
b-tagged jet permutation.          
  
In the second analysis the information of the b-tagger is not exploited. To 
increase the signal to background ratio several 
modifications of the selection cuts are applied. For example the transverse 
momenta of the first four jets are increased to 20 GeV. 
There are 87 $e$+jets and 80 $\mu$+jets  events left after this 
requirement. Next, using the specific kinematics of the 
$\ttbar$ events, a discriminant ($D$)  was constructed. It is 
designed to use variables which are uncorrelated or minimally  correlated 
with the  top quark mass\cite{D01998top}. Four topological variables 
are considered:

\begin{itemize}
\item $\met$ - missing transverse energy which comes from the neutrino of the W 
leptonic decay.
\item $\mathcal{A}$ - aplanarity of the event. It exploits the fact that the 
decay products from a massive particle have  large aplanarity. 
\item $H_{{T}^{2}}^{'}$ -  the ratio of the scalar sum of the 
$P_{T}$ of the jets, excluding the leading jet, 
and the scalar sum
of $|p_{z}|$ of the jets, the lepton and of the reconstructed neutrino.
\item $K_{{T}_min}^{'}$ - a measure of the jet separation folded 
with the $E_{T}$ of the reconstructed leptonic W boson. 
\end{itemize}  

Figure~4
shows the simulated discriminant distribution for   signal 
and background. A cut of $D>0.4$ is imposed to select the 
signal rich region. After the kinematical fit  at least one jet 
permutation is required to  have $\chi^{2}<$~10.    

\begin{figure}[t]
  \vspace{9.0cm}
  \includegraphics{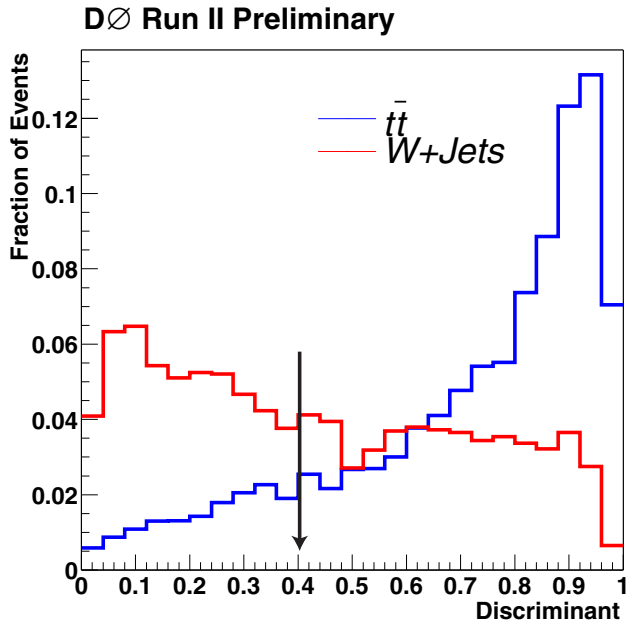}
  \caption{\it
 Discriminant D for $\ttbar$ (solid dark line) and background (solid light line) events from 
MC simulation.
    \label{fig:disc} }
\end{figure}


Similar to the CDF lepton plus jets analysis, two dominant sources  of 
background are accounted for:  
W plus multijet production, including heavy  flavour, and  QCD multijet events 
where one of 
the jets is misidentified as a lepton and there is  significant $\met$ imbalance 
in the event due to
detector resolution. 

The systematic uncertainties of both analyses are summarized in 
Table~\ref{tab:d0syst}. The  main contributions are due to 
JES, gluon radiation 
(initial state-ISR and final state-FSR), and the MC $\ttbar$ 
signal modeling. 

\begin{table}[t]
  \centering
\caption{Systematic uncertainties  in the
D\O~ lepton plus jets top quark mass measurement. 
The  uncertainty on JES, Gluon Radiation and Signal Model are the dominant sources 
of error on the mass.
    }
  \vskip 0.1 in
  \begin{tabular}{|c|c|c|} \hline
Source 	    	         	       &    \multicolumn{2}{c|} { $\Delta M_{top}$}        \\
                                       &    \multicolumn{2}{c|}  { (GeV/c$^{2}$) }           \\
                                       &  b-tagged analysis & Topological Analysis \\
    \hline
    \hline
  JES                                 &    +4.7/-5.3 & +6.8\-6.5  \\
  Jet Resolution                      &    $\pm$0.9 & $\pm$0.9 \\
  Gluon Radiation                     &    $\pm$2.4 & $\pm$2.6 \\
  Signal Model                        &    +2.3  &    +2.3 \\
  Background Model  	              &    $\pm$ 0.8 & $\pm$ 0.7 \\
  $b$-tagging	                      &    $\pm$0.7  & N/A \\
  Calibration                         &    $\pm$0.5  & $\pm$ 0.5 \\
  Trigger bias                        &    $\pm$0.5  & $\pm$ 0.5 \\
  MC statistics 	              &    $\pm$0.5  & $\pm$ 0.5 \\
  \hline				      	       					      				      
  Total			              &    $\pm6.0$ &   +7.8-7.1 \\
    \hline
  \end{tabular}
  \label{tab:d0syst}
\end{table}



As expected, the dominant systematic uncertainty originates from the JES. 
The  impact of JES on the  reconstructed top mass   was evaluated by 
scaling the jet energies by $\pm1\sigma$ for both signal and background 
in the MC simulation.
The uncertainty on  the JES was 
conservatively assumed  to be 5\% for the $E_{T}^{jet}>30$~GeV. For  jets 
with   $E_{T}^{jet}<30$~GeV, the 
JES uncertainty  decreases  linearly as  $\sigma = 
30\% - 25\% \times (E_{T}^{jet}/30$)~GeV.

Next in importance to the JES is the  systematic uncertainty coming from gluon 
radiation. Regardless of which jet 
permutation is used, the fitted mass will not be  correct  if the a radiated gluon 
is  one in the four leading  jets in the event. 
To understand how this  affects the $\ttbar$ 
reconstruction, MC  events with only four
partons hadronizing and forming four jets were compared to  events where 
one of the leading 4 jets comes from 
gluon radiation. 
A small deviation  of 
$\sim $ 0.2 GeV/c$^{2}$ from the nominal top mass is  observed when the 
events without gluon radiation are reconstructed. 
The difference becomes $\sim 2.4$  GeV/c$^{2}$ when 
one of the leading jets is a radiated gluon.
           
In this  analyses the model of the kinematic properties of the events is 
taken directly from MC simulation. Therefore some 
deficiencies in the MC model may lead to a substantial bias in the mass 
reconstruction. In order to perform a conservative 
estimate of this possible effect, in addition to the nominal sample for 
$\ttbar$ signal a complementary 
sample was generated where an additional parton is produced in 
association with the $\ttbar$ pair. The cross section
for this process is approximately two times smaller than the cross section 
for the $\ttbar$ production. By analyzing this sample, 
an uncertainty of +2.3~GeV/c${^2}$ due to the uncertainty on  signal modeling 
is assigned to the analyses. All other 
possible systematic effects turned out to be relatively small, at the 
level of 0.5$\sim$0.7 GeV/c$^{2}$.

The distributions of the fitted masses and  -ln($\mathcal{L}$) curves are 
shown in Figure~5.
The top 
two figures show the result from the $b$-tagged analysis while the bottom 
two represent the 
topological one. Taking into account the  output from the binned likelihood 
fit and the systematic uncertainties 
the final result for the analyses is $M_{top}$ = 170.6 $\pm4.2$(stat.) 
$\pm6.0$(syst.)~GeV/c$^{2}$ ($b$-tagged analysis) and 
 $M_{top}$ = 169.9$\pm5.8$(stat.)$^{+7.8}_{-7.1}$(syst.)~GeV/c$^{2}$ for 
the topological one.

\begin{figure}[t]
  \vspace{9.0cm}
  \includegraphics{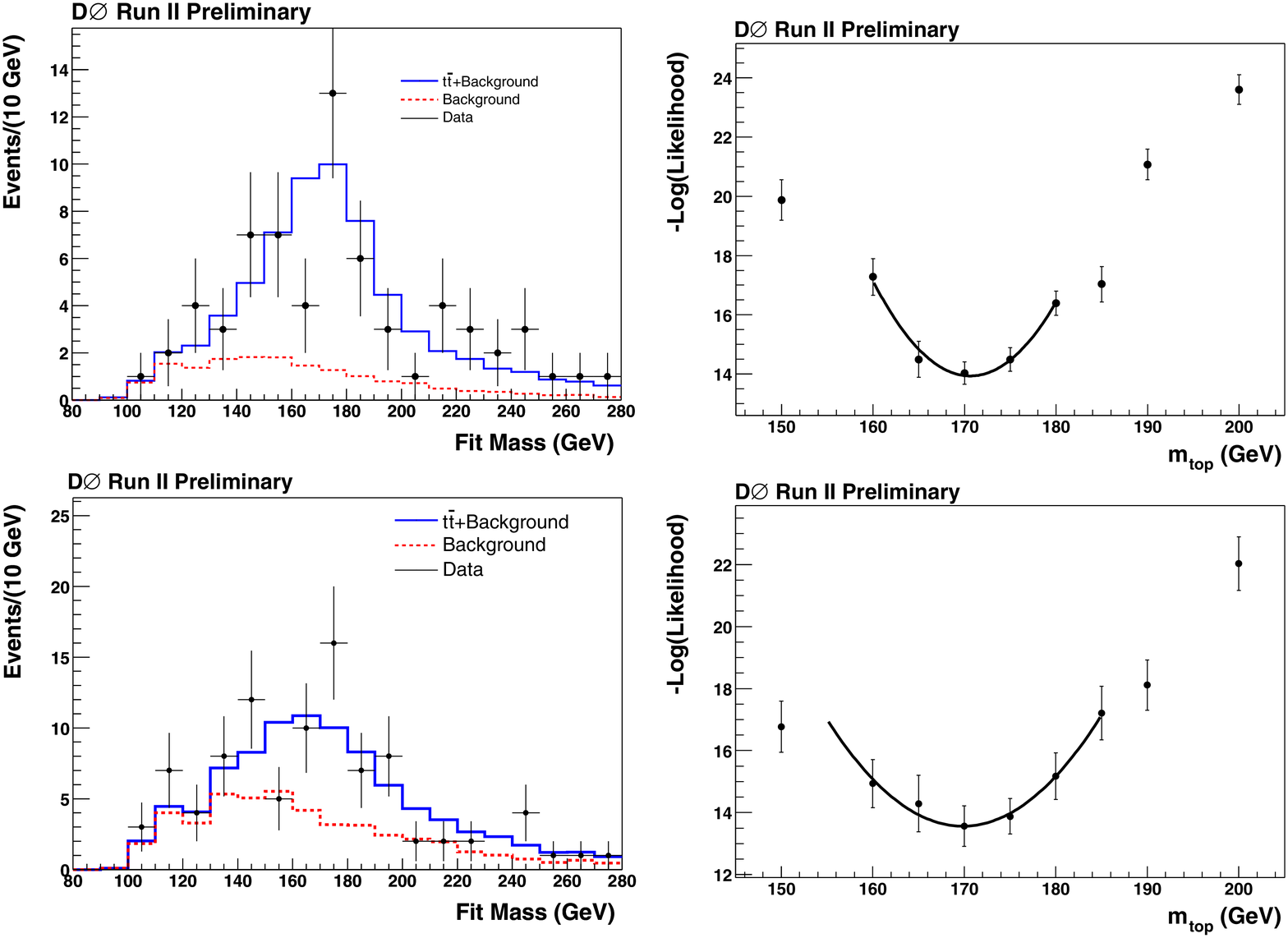}
  \caption{\it
The result from the binned likelihood fit for the $b$-tagged  
(topological) analysis  is presented on the upper (lower) left plot. 
The dots represent the data, 
the solid line is 
the fitted $\ttbar$ plus background and the dashed line is  background only, 
normalized to the fraction returned by the fit. The 
right plots show  the -ln($\mathcal{L}$) curves.
    \label{fig:D0final} }
\end{figure}


\section{ Dilepton Channel}
\subsection{ CDF result}
\hspace*{0.8cm} CDF has several independent dilepton 
analyses which are found to return consistent values for the top mass. 
Since this sample has good 
signal  to
background ratio ($\sim 4/1$) one is  stimulated to invent ingenious ways 
to reconstruct the events and extract $M_{top}$.

The event selection criteria are similar as in the lepton plus jets channel.
Two or more central jets with $\et>$ 15 GeV are required. A loose criterion is 
applied to the second lepton -
it must have  opposite charge but isolation is not mandatory.   
For the missing transverse energy the cut is increased to $\met>$25 GeV 
since two neutrinos are supposed to be presented in the event.
If $\met < $50 GeV, a  requirement for  the angle between $\met$
and the nearest lepton or jet to be $\Delta \phi>$20$^o$ is imposed.
Also the transverse energy sum, $H_{T}$, has to be more than 200~GeV.
Events due to cosmic rays, conversions or Z bosons are rejected.

Four major  backgrounds are taken into account: di-boson plus jet production, 
W plus jets where  one of the jet is 
faking a lepton, and Drell-Yan production, specially Z/$\gamma \rightarrow 
\tau\tau $. 33 events passed all
cuts with an  expected background of  11.6$\pm$2.1 events.    

In contrast to the lepton plus jets mode, 
in  the dilepton case due to the presence of two neutrinos the kinematics is not   
constrained.
The number of non-measured kinematical variables is larger by one  than the number 
of kinematic constraints 
($-1CF$). Obviously, it is impossible to single out only one solution 
per event. We may take  
some event parameter ($\vec{R}$) as known in order to constrain 
the kinematics and then 
vary  $\vec{R}$ to determine a set of solutions. 
In order to determine a preferred mass, every solution should have 
a weight attached to it.

The minimal requirement in the case of $-1CF$ kinematics is to use a two 
dimensional vector as $\vec{R}$.
We chose the azimuthal angles of the two neutrino momenta
$\vec{R}=(\phi_{\nu1},\phi_{\nu2})$ and create a net of solutions
in the $(\phi_{\nu1},\phi_{\nu2})$ plane.

For every point of the $(\phi_{\nu1},\phi_{\nu2})$ plane we have 8 solutions.
Two of them are generated by the two possibilities
of associating the two charged leptons to the two leading jets
which are assumed  to originate from the  $\bbbar$ partons.
The four other solutions are generated from the 
four ways of associating each neutrino to two  $p_z$
momenta,   satisfying the W decay  kinematics.  
We select the minimal $\chi^2$ 
solution for every point of the net for further use in our analysis.

Using the $\chi^2$ value from a minimization we weight the selected solutions 
by $e(-\frac{\chi^2}{2})$.
This is done in order to suppress the solutions which have worse compliance
with the fit hypothesis.

\begin{figure}[t]
\vspace{9.0cm}

  \includegraphics{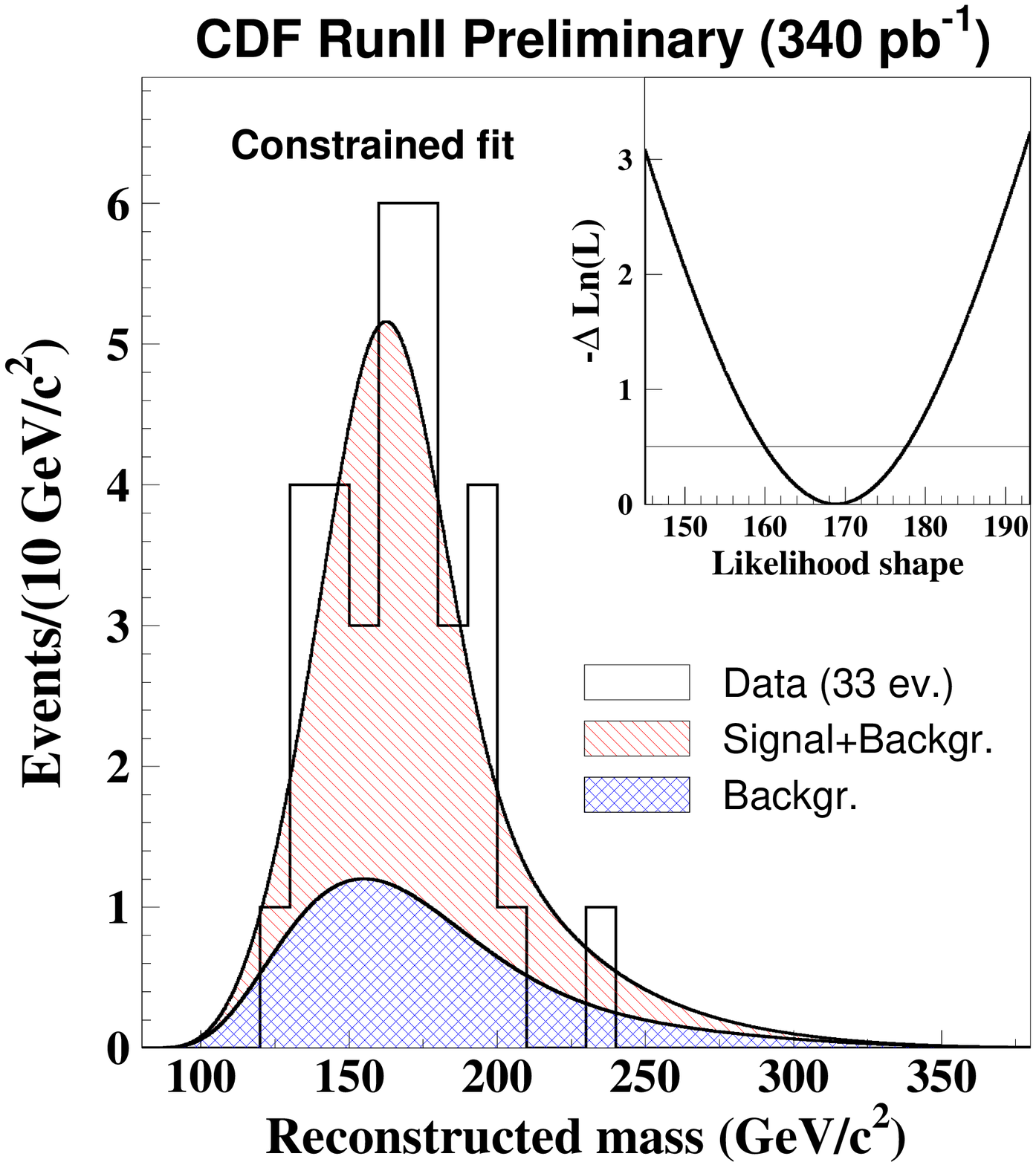}
   \includegraphics{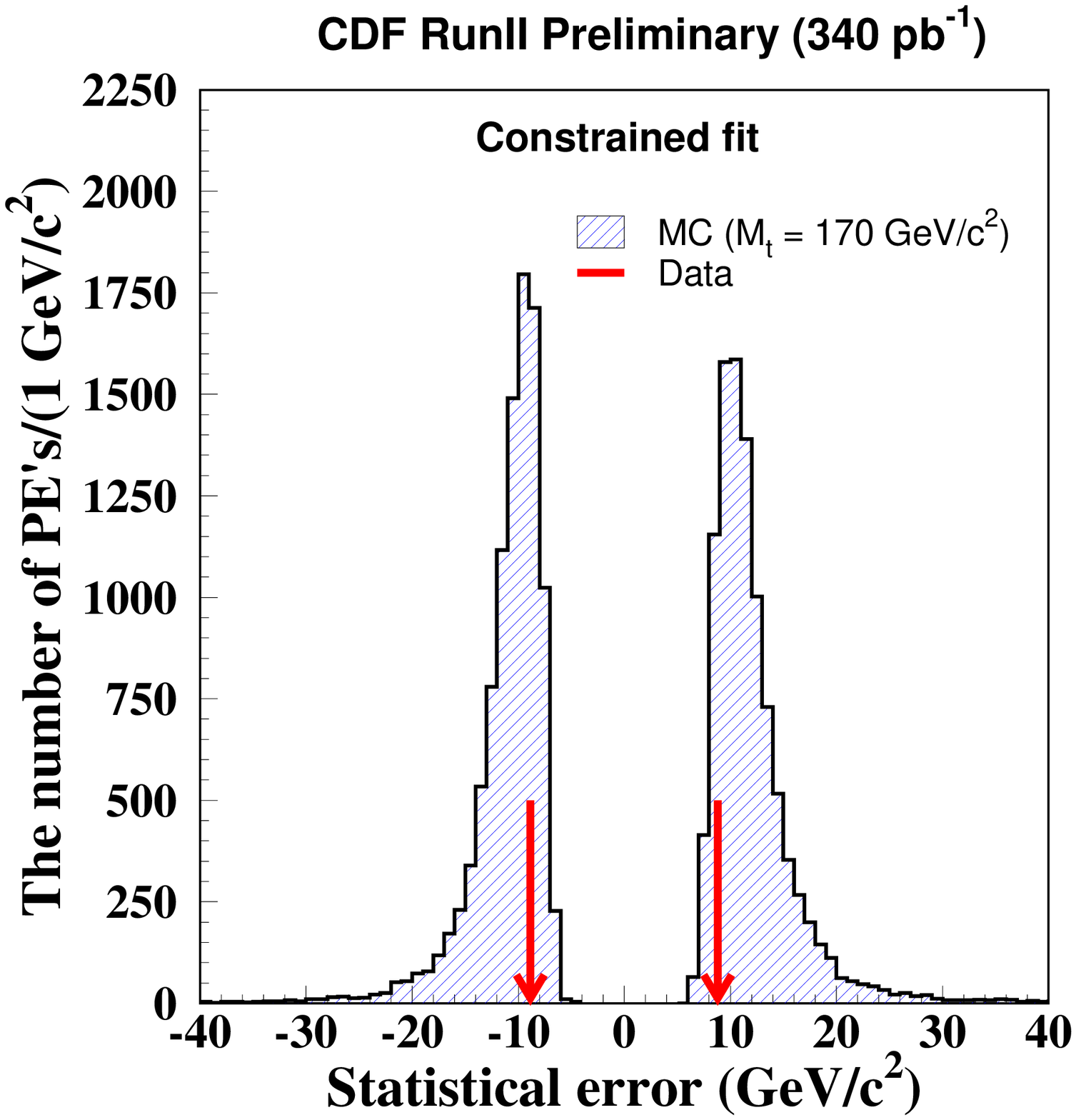} 
  \caption{\it
Left: two-component constrained fit to the dilepton sample. The cross hatched area
corresponds to the background returned by the fit and the 
line-shaded area is the sum of background
and signal.
The insert shows the mass-dependent negative log-likelihood used in the fit.
Right: positive and negative  error distributions returned by the fit in pseudo experiments. 
The arrows indicate the 
errors from the data fit.
    \label{fig:diltopmass} }
\end{figure}

The final extraction of the top quark mass from a sample of dilepton 
candidates is provided by
the unbinned likelihood fit. The expected signal and background p.d.fs are 
obtained
using Monte Carlo samples with detector simulation. The  background-constrained 
fit (N$_{b}$=11.6$\pm$2.1)  returns:
M$_{top}$ = 169.8 $\pm^{9.2}_{9.3}$ ~GeV/c$^2$,
with 23.4$\pm^{6.3}_{5.7}$ signal events.
The left plot in Fig.~\ref{fig:diltopmass} shows the fitted mass distribution. 
The insert shows the mass dependence
of the negative log-likelihood function.
The right plot represents  the error distribution 
for Monte-Carlo 
simulated experiments,
where the arrows indicate the data result.

We also performed a fit  without constraining  the number of  background events.
This fit returns
M$_{top}$ =  169.2$\pm^{6.4}_{6.5}$ ~GeV/c$^2$,
with 33.0$\pm^{6.0}_{5.8}$ signal events and
0.0$\pm^{4.2}_{0.0}$ background events.


\subsection{D\O~ result}
\hspace*{0.8cm} To reconstruct the top quark mass in the dilepton channel, D\O~ 
follows the ideas proposed by Dalitz and Goldstein in\cite{dalitzgoldstain}.
The analysis uses about 230~$\ipb$ of data. 
The  initial selection includes:
\begin{itemize}

\item two leptons, electron or muon, with $\pt>$ 15 GeV in the pseudorapidity 
regions  $|\eta(e)|<$1.1  or
1.5$<|\eta(e)|<$2.5 for the electron and  $|\eta(\mu)|<$2.0 for the muon. 
In e-$\mu$ events 
a separation cut of 
$\sqrt{\Delta\phi^{2}+\Delta\eta^{2}}>$0.25 is applied;
\item two or more jets with   $\pt>$ 20 GeV  in the pseudorapidity region 
$|\eta^{jet}|<$2.5;
\item large missing $\met >$25 GeV. However the $\met$ cut is varied for  
di-electron or di-muon events depending
on the $ee$ or $\mu\mu$ invariant mass;
\item veto on the $Z \rightarrow ee,\mu\mu$ events;
\item a cut $\Delta\phi(\mu,\met)>$ 0.25 rejects the events where 
the $\met$ and $\mu$ vectors are close to each other in the 
transverse plane;
\item $H_{T}>$ 140 GeV, where  $H_{T}$ is the scalar transverse momentum sum of the larger of the 
two lepton $\pt$s and of all jets over 15 GeV.  
\item for $ee$ events  a additional sphericity $>$ 0.15 cut is applied. 
\end{itemize}

8 $e\mu$, 5 $ee$ and 0 $\mu\mu$ events  satisfy all requirements, 
when 6.2$\pm$0.6, 2.8$\pm$0.3 and  2.9$\pm$0.6 events are 
correspondingly expected.  

The D\O~ analysis method can be 
summarized as follows. The momenta of the two highest $\pt$ jets in the event 
are assigned to the $\bbbar$ 
from the decay of $\ttbar$ quarks. Then a likelihood to hypothesized  values of 
the top  mass in the
region of 80$\sim$280~GeV is determined. For each event a solution is found 
when the pairs of $\ttbar$ momenta are 
consistent with the observed lepton and jet momenta 
and $\met$.
A weight to each solution is assigned as

\begin{equation}
\mathcal{W}= f(x) f(\bar{x}) p(E_{\ell}^{*}|M_{top})  p(E_{\bar{\ell}}^{*}|M_{top}), 
\label{lhood2}
\end{equation}    

where $f(x)$ ($f(\bar{x})$) is the parton distribution function for the proton 
(anti-proton) and the initial quark (anti-quark) is carrying 
a momentum fraction $x$ ($ \bar{{\rm x}}$). $p(E_{\ell}^{*}|M_{top})$ denotes the probability for 
the top (anti-top) quark with a mass  $M_{t}$ to generate 
a lepton $\ell$ ($\bar{\ell}$) with the observed energy in the top quark rest 
frame. 

There are two ways to assign the two jets to the $b$ and $\bar{b}$ 
quarks. In addition, for each jet-to-parton assignment,
there might be up to four solutions for each hypothesized value of the mass, 
coming from the fact that every neutrino may have up to 
two real solutions for its $p_{z}$, satisfying the  kinematics. 
Then the likelihood for each value of the top quark mass $M_{top}$ is given 
by the sum of the weights $w_{i,j}$ over all possible solutions:

\begin{equation}
\mathcal{W}(M_{t})= \sum_{p_{z}^{\nu}~solutions}^{} \sum_{jet~assignment}^{} w_{i,j}.
\label{lhood3}
\end{equation}   

\begin{figure}[t]
  \vspace{9.0cm}
  \includegraphics{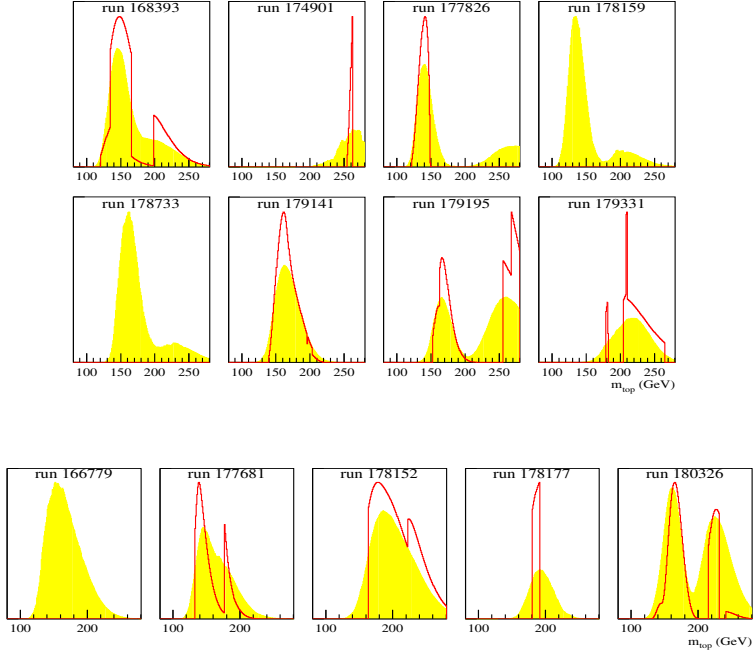}
  \caption{\it
Weight curves from  8 $e\mu$  events (top) and 5 $ee$ events (bottom). 
The shadow histograms show the weight curves with 
resolution smearing  while the open histograms represent the weight curves 
without resolution smearing.
    \label{fig:d0dilep} }
\end{figure}

Up to now there was an implicit assumption that  all  momenta are measured 
perfectly. Therefore the weight in (\ref{lhood3}) is zero
if no exact solutions are found. To account for  detector resolution
 the weight calculations are  repeated with input values for 
the particle momenta   drawn from  Gaussian distributions with  
means equal of the  measured values and widths corresponding to 
the  detector resolution. In addition the $\met$ value is recalculated  
from generated particle momenta and a random noise from 
a normal distribution with mean  0 GeV and rms  8 GeV is added. 
Figure~\ref{fig:d0dilep} up (down) shows the weight curves for $e\mu$ 
($ee$) events before (solid line)  and after (shadow area) resolution 
smearing. For each event the value of the  top
quark mass at which the weight curve reaches its  maximum is used as the 
estimator of the mass. After that, to extract 
the most probable top mass value from the data sample,  a standard 
template method which exploits a binned maximum likelihood fit is applied. 
The  likelihood fit returns  
$M_{top} $ =  $155.^{+14}_{-13} (stat.) \pm 7. (syst.)$  GeV/c$^{2}$. The 
JES uncertainty (5.6 GeV/c$^2$) dominates  the systematic error.  

\newpage
\section{Summary of the Top Quark Mass Measurements and Run~II Prospects}
\hspace*{0.8cm}

Combining the presently available most accurate  
Run~II  CDF and D\O~ measurements
in the    dilepton and 
lepton plus jets decay topologies,
one finds $172.7\pm3.5$~GeV/c${^2}$. This  result is unofficial. 
The average  is made by the author assuming   
simple correlations (0 or 1)  between  the systematic
uncertainties in the  CDF and D\O~
measurements.  

The expected CDF uncertainty for JES systematics as a function of integrated luminosity 
is shown in 
Figure~\ref{fig:summary}, left.
The  right plots shows the total top mass error versus integrated 
luminosity for the 
CDF lepton plus jet analysis. One may  conclude that with a Run~II integrated 
luminosity of 8~$\ifb$
the top quark mass could be measured by CDF with a precision of  $\sim$2.0 GeV/c$^{2}$. 
This optimistic forecast is based on the present understanding  
that both the statistical and JES systematic uncertainties 
will decrease as expected with increasing integrated luminosity. 

\begin{figure}[t]
  \vspace{9.0cm}
  \includegraphics{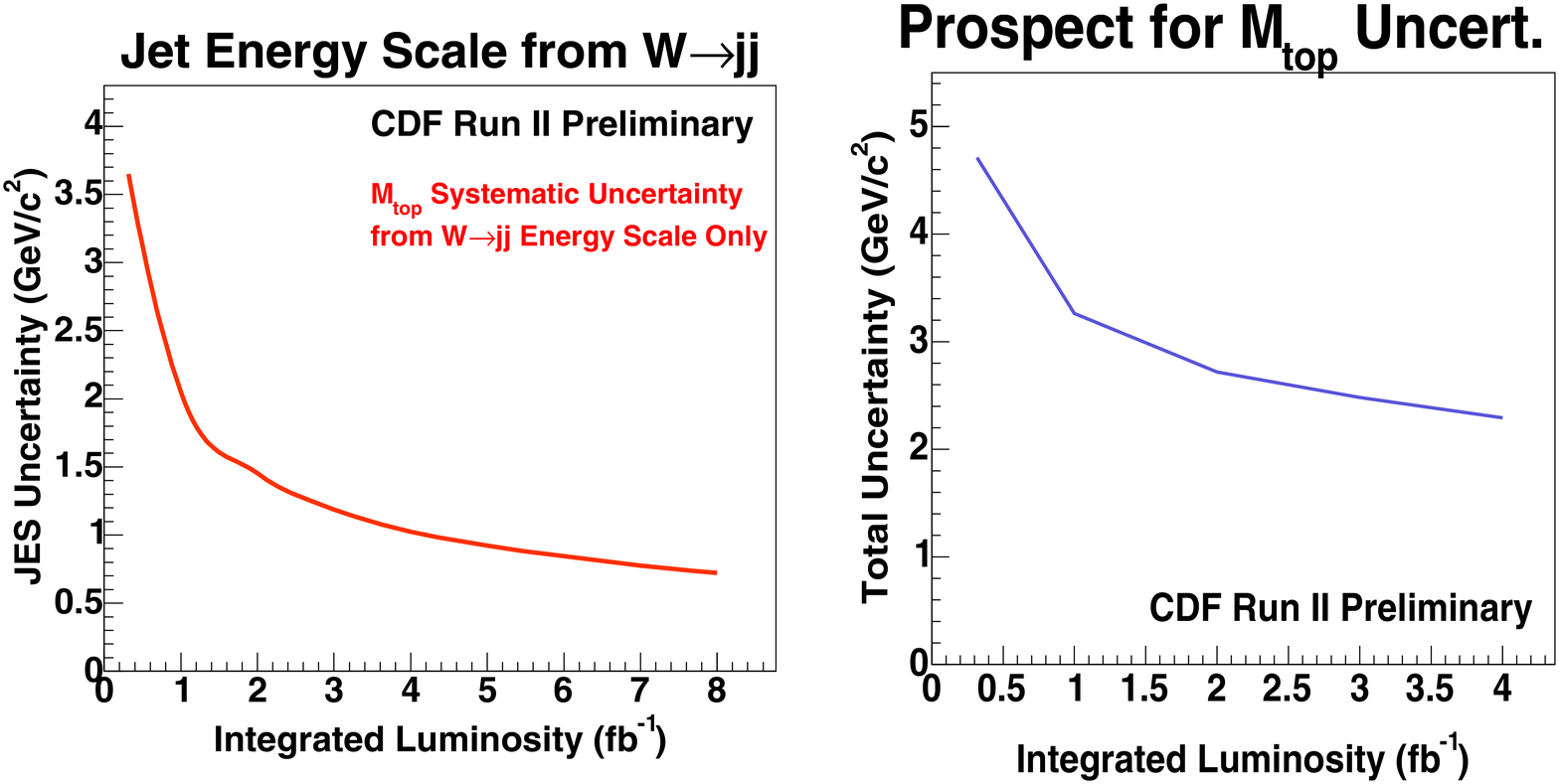}
  \caption{\it
 The expected JES  uncertainty from W$\rightarrow$jj  as a function of integrated 
luminosity is  shown on the left plot. On the right,  the total expected top mass uncertainty, 
from CDF lepton plus jets events   as a function of integrated luminosity, is shown.
    \label{fig:summary} }
\end{figure}


\section{Conclusion}
\hspace*{0.8cm}
The top quark CDF results from the Tevatron 2002-2004 Run~II,  with an
integrated luminosity of 318  and 230  ~$\ipb$ for CDF and D\O~ are
presented. 
The best, up to date, measurement of the top quark mass from the CDF lepton plus 
jets  analysis is 173.5$^{+3.9}_{-3.8}$ ~GeV/c$^{2}$.
Combining the CDF and D\O~ dilepton and  lepton plus jets Run~II results, the 
author's average  of the top quark mass is    $172.7\pm3.5$~GeV/c${^2}$.

\section{Acknowledgments}
\hspace*{0.8cm} I would like to thank the organizers for their
invitation and hospitality. This work is supported by the U.S. 
department of Energy and CDF and D~O\ collaborating institutions  and 
their funding agencies.

\end{document}